\documentclass[conference]{IEEEtran}
\IEEEoverridecommandlockouts

\usepackage{adjustbox}
\usepackage{amsmath,amssymb,amsfonts}
\usepackage{arydshln}
\usepackage{bm}  
\usepackage{booktabs}
\usepackage{cite}
\usepackage{color, soul}
\usepackage{enumitem}
\usepackage{multirow}
\usepackage{setspace}
\usepackage{subfigure}
\usepackage[normalem]{ulem}
\usepackage{xspace}

\usepackage[dvipsnames]{xcolor}
\usepackage[colorlinks=true,citecolor=black,linkcolor=black, urlcolor=black]{hyperref}
\usepackage{cleveref}
\usepackage[]{collab}

\usepackage[most]{tcolorbox}
\definecolor{background}{HTML}{D9E7FF}
\definecolor{edge}{HTML}{8686DD}
\newtcolorbox{mybox}{colback=background!30,%gray background
colframe=edge,% black frame colour
width=\columnwidth,% total width
boxrule = 0.3mm,
top = 3pt, bottom=3pt, left=3pt, right=3pt
}

\newcommand{\name}{{Eadro}\xspace}
\def\A{$\mathcal{TT}$\xspace}
\def\B{$\mathcal{SN}$\xspace}

\crefname{section}{§}{§§}
\Crefname{section}{§}{§§}
\begin{document}
\title{\name: An End-to-End Troubleshooting Framework for Microservices on Multi-source Data}

\author{
  \IEEEauthorblockN{
    Cheryl Lee\IEEEauthorrefmark{1},
    Tianyi Yang\IEEEauthorrefmark{1},
    Zhuangbin Chen\IEEEauthorrefmark{1},
    Yuxin Su\IEEEauthorrefmark{2}\thanks{Yuxin Su is the corresponding author.}, 
    and
    Michael R. Lyu\IEEEauthorrefmark{1}
  }

  \IEEEauthorblockA{\IEEEauthorrefmark{1}Department of Computer Science and Engineering, The Chinese University of Hong Kong, Hong Kong, China.\\
    Email: cheryllee@link.cuhk.edu.hk, \{tyyang, zbchen, lyu\}@cse.cuhk.edu.hk}

  \IEEEauthorblockA{\IEEEauthorrefmark{2}Sun Yat-sen University, Guangzhou, China.
    Email: suyx35@mail.sysu.edu.cn}
}

\maketitle
\begin{abstract}
The complexity and dynamism of microservices pose significant challenges to system reliability, and thereby, automated troubleshooting is crucial.
Effective root cause localization after anomaly detection is crucial for ensuring the reliability of microservice systems.
However, two significant issues rest in existing approaches: 
(1) Microservices generate traces, system logs, and key performance indicators (KPIs), but existing approaches usually consider traces only, failing to understand the system fully as traces cannot depict all anomalies; 
(2) Troubleshooting microservices generally contains two main phases, i.e., anomaly detection and root cause localization. 
Existing studies regard these two phases as independent, ignoring their close correlation. Even worse, inaccurate detection results can deeply affect localization effectiveness.
To overcome these limitations, we propose \textit{Eadro}, the first end-to-end framework to integrate anomaly detection and root cause localization based on multi-source data for troubleshooting large-scale microservices.
The key insights of Eadro are the anomaly manifestations on different data sources and the close connection between detection and localization.
Thus, Eadro models intra-service behaviors and inter-service dependencies from traces, logs, and KPIs, all the while leveraging the shared knowledge of the two phases via multi-task learning.
Experiments on two widely-used benchmark microservices demonstrate that \name outperforms state-of-the-art approaches by a large margin. The results also show the usefulness of integrating multi-source data.
We also release our code and data to facilitate future research.
\end{abstract}

\section{Introduction}\label{sec:intro}
\begin{figure}[htb]
    \centering
        {\includegraphics[width=0.65\linewidth]{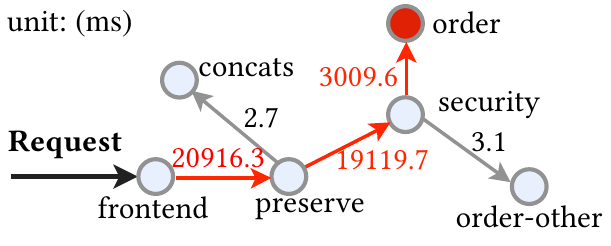}}
    \caption{A failure in ``order'' indirectly delays other microservices on the invocation chain, while microservices off the chain are unaffected.}
    \label{fig:dependency-graph}
\end{figure}

Microservice systems are increasingly appealing to cloud-native enterprise applications for several reasons, including resource flexibility, loosely-coupled architecture, and lightweight deployment~\cite{Ali21Survey}.
However, anomalies are inevitable in microservices due to their complexity and dynamism. An anomaly in one microservice could propagate to others and magnify its impact, resulting in considerable revenue and reputation loss for companies~\cite{IndSurvey21Zhou}.
Figure~\ref{fig:dependency-graph} shows an example where a failure in one microservice may delay all microservices on the invocation chain.

Therefore, developers must closely monitor the microservice status via run-time information (e.g., traces, system logs, and KPIs) to discover and tackle potential failures in their earliest efforts.
Yet, thousands of microservices are usually running in distributed machines in a large-scale industrial microservice system. As each microservice can launch multiple instances, a system can produce billions of run-time records per day~\cite{IndSurvey21Zhou, Ali21Survey}. The explosion of monitoring data makes automated troubleshooting techniques imperative.

Many efforts have been devoted to this end, focusing either on anomaly detection~\cite{TraceAnomaly, MMTrace, DeepTraLog} or on root cause localization~\cite{DyCause, MicroCause, MonitorRank, MicroHECL, TBAC, CloudRanger}. Anomaly detection tells whether an anomaly exists, and root cause localization identifies the culprit microservice upon the existence of an anomaly.
Previous approaches usually leverage statistical models or machine learning techniques to mine information from traces, as traces profile and monitor microservice executions and record essential inter-service information (e.g., request duration).
However, we identify two main limitations of the existing troubleshooting approaches.

(1) \textit{Insufficient exploitation of monitoring data}:
different from operation teams that pay close attention to diverse sources of run-time information, existing research deeply relies on traces and exploits other data sources insufficiently. 
This gap stems from the complexity of multi-source data analysis, which is much harder than single-source data analysis, as multi-source data is heterogeneous, frequently interacting, and very large~\cite{Multisource}.
However, on the one hand, traces contain important information for troubleshooting but are insufficient to reveal all typical types of anomalies.
On the other hand, different types of data, such as logs and KPIs, can reveal anomalies collaboratively and bring more clues about potential failures.
For example, a CPU exhaustion fault can cause abnormally high values in the CPU usage indicator and trigger warnings recorded in logs, but the traces may not exhibit abnormal patterns (such as high latency).

(2) \textit{Disconnection in closely related tasks}:
Generally, root cause localization follows anomaly detection since we must discover an anomaly before analyzing it.
Current studies of microservice reliability regard the two phases as independent, despite their shared inputs and knowledge about the microservice status. Existing approaches usually deal with the same inputs redundantly and waste the rich correlation information between anomaly detection and root cause localization.
Moreover, the contradiction between computing efficiency and accuracy limits the simple combination of state-of-the-art anomaly detectors and root cause localizers.
For a two-stage troubleshooting approach, it is generally a little late to use an advanced anomaly detector and then analyze the root cause.
Thus, root cause localization-focused studies usually apply oversimplified anomaly detectors (e.g., N-sigma), and unfortunately, the resulting detection outputs can contain many noisy labels and thereby affect the effectiveness of downstream root cause localization. 

To overcome the above limitations, we propose \textbf{\name}, the first \underline{E}nd-to-end framework integrating \underline{A}nomaly \underline{D}etection and \underline{R}oot cause l\underline{O}calization to troubleshoot microservice systems based on multi-source monitoring data.
The key ideas are 1) learning discriminative representations of the microservice status via multi-modal learning and 2) forcing the model to learn fundamental features revealing anomalies via multi-task learning.
Therefore, \name can fully exploit meaningful information from different data sources that can all manifest anomalies. Also, it allows information to be inputted once and used to tackle anomaly detection and root cause localization together and avoids incorrect detection results hindering next-phase root cause localization.

Specifically, \name consists of three components:
\textit{(1) Modal-wise learning} contains modality-specific modules for learning intra-service behaviors from logs, KPIs, and traces. 
We apply Hawkes process~\cite{Hawkes} and a fully connected (FC) layer to model the log event occurrences.
KPIs are fed into a dilated causal convolution (DCC) layer~\cite{TCN} to learn temporal dependencies and inter-series associations.
We also use DCC to capture meaningful fluctuations of latency in traces, such as extremely high values.
\textit{(2) Dependency-aware status learning} aims to model the intra- and inter-dependencies between microservices.
It first fuses the multi-modal representations via gated concentration and feeds the fused representation into a graph attention network (GAT), where the topological dependency is built on historical invocations.
\textit{(3) Joint detection and localization} contains an anomaly detector and a root cause localizer sharing representations and an objective. It predicts the existence of anomalies and the probability of each microservice being the culprit upon an anomaly alarm.

Experimental results on two datasets collected from two widely-used benchmark microservice systems demonstrate the effectiveness of \name. 
For anomaly detection, \name surpasses all compared approaches by a large margin 
in \textit{F1} (53.82\%\textasciitilde92.68\%), and also increases \textit{F1} by 11.47\% on average compared to our derived multi-source data-based methods.
For root cause localization, \name achieves state-of-the-art results with 290\%\textasciitilde5068\% higher in \textit{HR@1} (Top-1 Hit Rate) than five advanced baselines and outperforms our derived methods by 43.06\% in \textit{HR@1} on average.
An extensive ablation study further confirms the contributions of modeling different data sources.

Our main contributions are highlighted as follows:
\begin{itemize}%[leftmargin=10pt, topsep=0pt]
\setlength\itemsep{0em}
   
   \item We identify two limitations of existing approaches for troubleshooting microservices, motivated by which we are the first to explore the opportunity and necessity to integrate anomaly detection and root cause localization, as well as exploit logs, KPIs, and traces together.
    
    \item We propose the first end-to-end troubleshooting framework (\name) to jointly conduct anomaly detection and root cause localization for microservices based on multi-source data. \name models intra-service behaviors and inter-service dependencies.
    
    \item We conduct extensive experiments on two benchmark datasets. The results demonstrate that \name outperforms all compared approaches, including state-of-the-art approaches and derived multi-source baselines on both anomaly detection and root cause localization. We also conduct ablation studies to further validate the contributions of different data sources. 
    
    \item Our code and data~\footnote{https://github.com/BEbillionaireUSD/Eadro} are made public for practitioners to adopt, replicate or extend \name.
    
\end{itemize}

\section{Problem Statement}
This section introduces important terminologies and defines the problem of integrating anomaly detection and root cause localization with the same inputs.

\subsection{Terminologies}
Traces record the process of the microservice system responding to a user request (e.g., click ``create an order'' on an online shopping website). 
Different microservice instances then conduct a series of small actions to respond to the request.
For example, the request ``create an order'' may contain steps ``create an order in pending'', ``reserve credit'', and ``update the order state.''
A microservice (caller) can \textit{invoke} another microservice (callee) to conduct the following action (e.g., microservice ``Query'' asks microservice ``Check'' to check the order after finishing the action ``query the stock of goods''), and the callee will return the result of the action to the caller. We name this process as \textit{invocation}.
The time consumed by the whole invocation (i.e., from initializing the invocation to returning the result) is called invocation \textit{latency}, including the request processing time inside a microservice and the time spent on communicating between the caller and the callee. 
A \textit{trace} records the information during processing a user request~\cite{MEPFL}  (including multiple invocations), such as the invocation latency, the total time of processing the request, the HTTP response code, etc.

Meanwhile, system logs are generated when system events are triggered. 
A \textit{log message} (or \textit{log} for short) is a line of the standard output of logging statements, composed of constant strings (written by developers) and variable values (determined by the system)~\cite{SurveyHe}. If the variable values are removed, the remaining constant strings constitute a \textit{log event}. 
\textit{KPIs} are the numerical measurements of system performance (e.g., disk I/O rate) and the usage of resources (e.g., CPU, memory, disk) that are sampled uniformly. 

\subsection{Problem Formulation}\label{sec:prob form}
Consider a large-scale system with $M$ microservices, system logs, KPIs, and traces are aggregated individually at each microservice.
In a $T$-length observation window (data obtained in a window constitute a sample), we have multi-source data defined as $\mathbf{X}=\{ ( \mathbf{X}^\mathcal{L}_{m}, \mathbf{X}^\mathcal{K}_{m}, \mathbf{X}^\mathcal{T}_{m} ) \}_{m=1}^M$, 
where at the $m$-th microservice,
$\mathbf{X}^\mathcal{L}_{m}$ represents the log events chronologically arranged;
$\mathbf{X}^\mathcal{K}_{m}$ is a multivariate time series consisting of $k$ indicators;
$\mathbf{X}^\mathcal{T}_{m}$ denotes the trace records.
Our work attempts to build an end-to-end framework achieving a two-stage goal:
Given $\mathbf{X}_{[1:M]}$, the framework predicts the existence of anomalies, denoted by $y$, a binary indicator represented as 0 (normal) or 1 (abnormal). 
If $y$ equals one, a localizer is triggered to estimate the probability of each microservice to be the culprit, denoted by $\textbf{P}=[p_{1} \cdots p_{M}] \in [0, 1]^M$.
The framework is built on a parameterized model $\mathcal{F}: \mathbf{X} \rightarrow (y, \mathbf{P})$.

\section{Motivation}\label{sec:motivation}
This section introduces the motivation for this work, which aims to address effective root cause localization by jointly integrating an accurate anomaly detector and being driven by multi-source monitoring data.
The examples are taken from data collected from a benchmark microservice system, TrainTicket~\cite{TrainTicket}. Details about data collection will be introduced in~\cref{sec:data}.

\subsection{Can different sources of data besides traces be helpful?}\label{sec:study:multi-source}
\begin{figure}[htb]
    \centering
    \vspace{-0.1in}
        \includegraphics[width=0.97\linewidth]{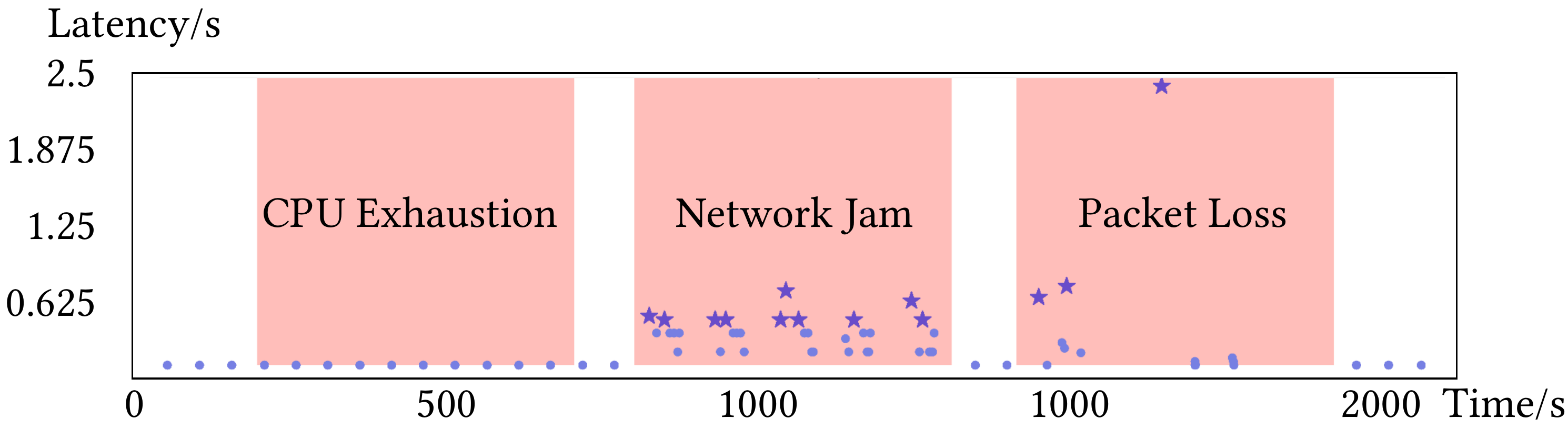}
    \caption{Network-related faults incur obvious anomalies in latency of ``travel'', but the CPU exhaustion fault does not.}
    \label{fig:travel-latency}
\vspace{-0.1in}
\end{figure}

We find that \textit{traces are insufficient to reveal all potential faults despite their wide usage.}
Most, if not all, previous related works~\cite{Trace17, TraceAnomaly, TraceRCA, MMTrace, MicroCause, MicroScope, MicroRank} are trace-based, indicating traces are informative and valuable.
However, traces focus on recording interactions between microservices and provide a holistic view of the system in practice.
Such high-level information only enables basic queries for coarse-grained information rather than intra-service information. 
For example, latency or error rate in traces can suggest a microservice's availability, yet fine-grained information like memory usage reflecting the intra-service status is unknowable.
%%%%% Latency is useful but not sufficient
This is consistent with our observation that latency is sensitive to network-related issues but cannot adequately reflect resource exhaustion-related anomalies.
Figure~\ref{fig:travel-latency} shows an example where a point denotes an invocation taking the microservice ``travel'' as the callee.
When Network Jam or Packet Loss is injected, the latency is abnormally high (marked with stars), but the latency during the CPU exhaustion injection period does not display obviously abnormal patterns. 
This case reminds us to be careful of relying on traces only. 
Since traces are informative but cannot reveal all anomalies, trace-based methods may omit potential failures. We need extra information to mitigate the anomaly omission problem.

We also notice that \textit{system logs and KPIs provide valuable information manifesting anomalies in microservices.}

As for logs, we first parse all logs into events via Drain~\cite{Drain}, a popular log parser showing effectiveness in many studies~\cite{SurveyHe, reportZB}.
It is evident that some logs can report anomalies semantically by including keywords such as ``exception'', ``fail'', and ``errors''. The event ``Exception in monitor thread while connecting to server $<$\textit{*}$>$.'' can be a good example. 

\begin{figure}[htb]
    \centering
    \vspace{-0.1in}
    \includegraphics[width=0.95\linewidth]{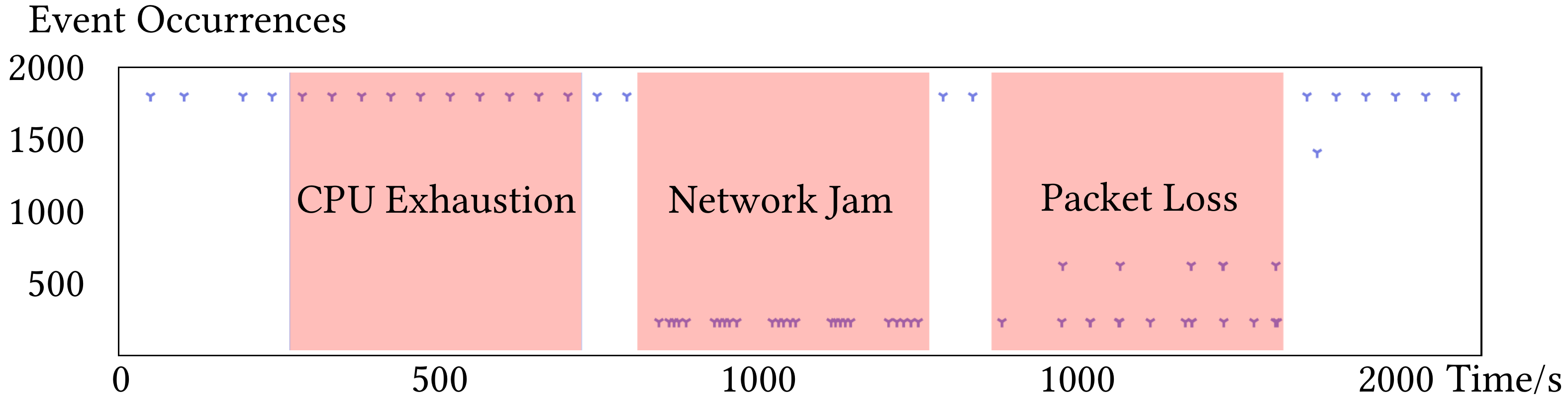}
    \caption{The occurrences of related logs can reflect issues such as poor communication.}
    \label{fig:route-log}
\vspace{-0.1in}
\end{figure}

Event occurrences can also manifest anomalies besides semantics. 
Take the event ``Route id: $<$\textit{*}$>$'' recorded by the microservice ``route'' as an example. This event occurs when the microservice completes the routing request.
Figure~\ref{fig:route-log} shows that when network-related faults are injected, the example event's occurrence experiences a sudden drop and remains at low values. The reason is that the routing invocations become less since the communication between ``route'' and its parent microservices (callers) is blocked.
This case further supports our intuition that system logs can provide clues about microservice anomalies.

\begin{figure}[htb]
    \centering
    \vspace{-0.1in}
    \includegraphics[width=0.95\linewidth]{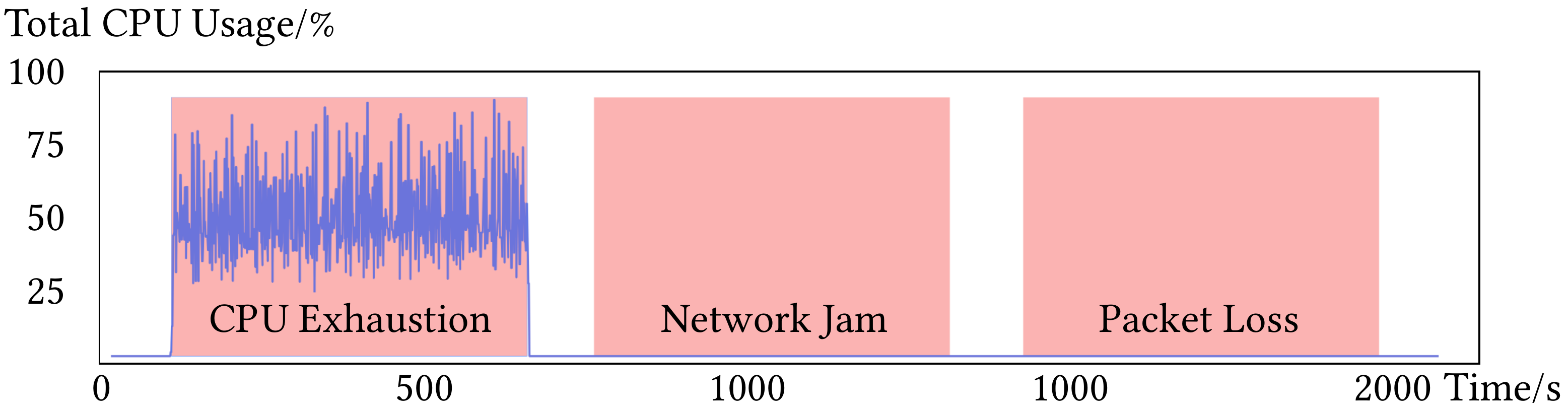}
    \caption{A CPU exhaustion fault incurs abnormal jitters and high values in ``total CPU usage''.}
    \label{fig:payment}
\vspace{-0.1in}
\end{figure}

KPIs are responsive to anomalies by continuously recording run-time information.
An example in Figure~\ref{fig:payment} gives a closer look, which displays ``total CPU usage'' of microservice ``payment'' during the period covering fault injections.
Clearly, ``total CPU usage'' responds to the fault CPU Exhaustion by showing irregular jitters and abnormally high values. This observation aligns with our a priori knowledge that KPIs provide an external view of a microservice's resource usage and performance. Their fine-grained information can well reflect anomalies, especially resource-related issues, which require detailed analysis.

However, only using logs and KPIs is not sufficient since they are generated by each microservice individually at a local level. As the example shown in Figure~\ref{fig:dependency-graph} (\cref{sec:intro}), we need traces to obtain inter-service dependencies to analyze the anomaly propagation so as to draw a global picture of the system to locate the root cause.

\begin{mybox}
    Traces are informative yet not sufficient to reflect all anomalies. System logs and metrics provide valuable information manifesting anomalies by presenting abnormal patterns, so they can serve as additional information. 
\end{mybox}

\subsection{Can current anomaly detectors provide accurate results?}\label{sec:study:detection}
This section demonstrates that \textit{current detectors attached with localizers cannot deliver satisfying accuracy}. 

As far as we know, existing root cause localization approaches for microservices follow such a pipeline: 1) conduct anomaly detection and 2) if an anomaly is alarmed, then the localizer is triggered. That is, the anomaly detector and root cause localizer work separately.
Unfortunately, incorrect anomaly detection  results can exert a negative impact on the following root cause localization by introducing noisy labels.
To investigate whether current anomaly detectors are satisfactory for downstream localizers, we first summarize three main kinds of anomaly detection  approaches used in root cause localization papers. Note that since this paper targets root cause localization, the listed approaches are root cause localization-oriented anomaly detectors rather than sophisticated approaches for general anomaly detection.

\begin{itemize}[leftmargin=12pt, topsep=0pt]
\item N-sigma used in~\cite{MicroRank, MicroScope} computes the mean ($\mu$) and the standard deviation ($\sigma$) of historical fault-free data. If the maximum latency of the current observation window is larger than $\mu+n\cdot \sigma$, an alarm will be triggered, where $n$ is an empirical parameter.

\item Feature engineering + machine learning (FE+ML)~\cite{MicroHECL, AutoMAP} feeds manually derived features from traces into a machine learning-based model such as OC-SVM~\cite{MicroHECL} to detect anomalies in a one-class-classification manner. 

% The features are listed in Table~\ref{tab:features}.
\item SPOT~\cite{SPOT} is an advanced algorithm for time series anomaly detection based on the Extreme Value Theory. Recent root cause analysis studies~\cite{DyCause, MicroCause} have applied it for detecting anomalies.

\end{itemize}
\begin{table}[htbp]
\vspace{-0.1in}
\small
\centering
\caption{Comparison of Common Anomaly Detectors}
\begin{tabular}{cccc}
\toprule
& \textbf{N-sigma} & \textbf{FE+ML} & \textbf{SPOT} \\
\midrule
FOR & 0.632 & 0.830 & 0.638\\
FDR & 0.418 & 0.095 & 0\\
\#Infer/ms & 0.207 & 1.361 & 549.169\\
\bottomrule
\end{tabular}
\vspace{-0.1in}
\label{tab:fault-detection}
\end{table}

We conduct effectiveness measurement experiments based on our data on the three anomaly detectors following~\cite{MicroRank, MicroHECL, DyCause}, respectively. We focus on the false omission rate (\textit{FOR}$=\frac{FN}{FN+TN}$) and the false discovery rate (\textit{FDR}$=\frac{FP}{FP+TN}$), where
$TN$ is the number of successfully predicted normal samples; 
$FN$ is the number of undetected anomalies;
$FP$ is the number of normal samples incorrectly triggering alarms.
Besides, \#Infer/ms denotes the average inference time with the unit of microseconds.

Table~\ref{tab:fault-detection} lists the experimental results, demonstrating a large improvement space for these anomaly detectors. The high \textit{FOR} and \textit{FDR} indicate that the inputs of the root cause localizer contain lots of noisy labels, thereby substantially influencing localization performance.
We attribute this partly to the closed-world assumption relied on by these methods, that is, regarding normal but unseen data patterns as abnormal, thereby incorrectly forcing the downstream localizer to search for the ``inexistent'' root cause based on normal data.
Also, latency is insufficient to reveal all anomalies, as stated before, especially those that do not severely delay inter-service communications, represented by the high \textit{FOR}. 

In addition, complex methods (FE+ML and SPOT) have better effectiveness than N-sigma yet burden the troubleshooting process by introducing extra computation.
Since root cause localization requires anomaly detection first, the detector must be lightweight to mitigate the efficiency reduction. 
Even worse, these machine learning-based approaches require extra hyperparameter tuning, making the entire troubleshooting approach less practical.
\vspace{0.05in}
\begin{mybox}
    Root cause localization requires anomalous data detected by anomaly detectors, but current localization-oriented detectors either deliver unsatisfactory accuracy and introduce noisy data or reduce efficiency, making the following localization troublesome. 
\end{mybox}

In summary, these examples motivate us to design an end-to-end framework that integrates effective anomaly detection and root cause localization in microservices based on multi-source information, i.e., logs, KPIs, and traces.
Logs, KPIs, and latency in traces provide local information on intra-service behaviors, while invocation chains recorded in traces depict the interactions between microservices, thereby providing a global view of the system status.
This results in \name, the first work to enable jointly detecting anomalies and locating the root cause, all the while attacking the above-mentioned limitations by learning the microservice status concerning both intra- and inter-service properties from various types of data.

\section{Methodology}\label{sec:method} 
\begin{figure*}[htb]
    \centering
    \vspace{-0.2in}
        {\includegraphics[width=1\linewidth]{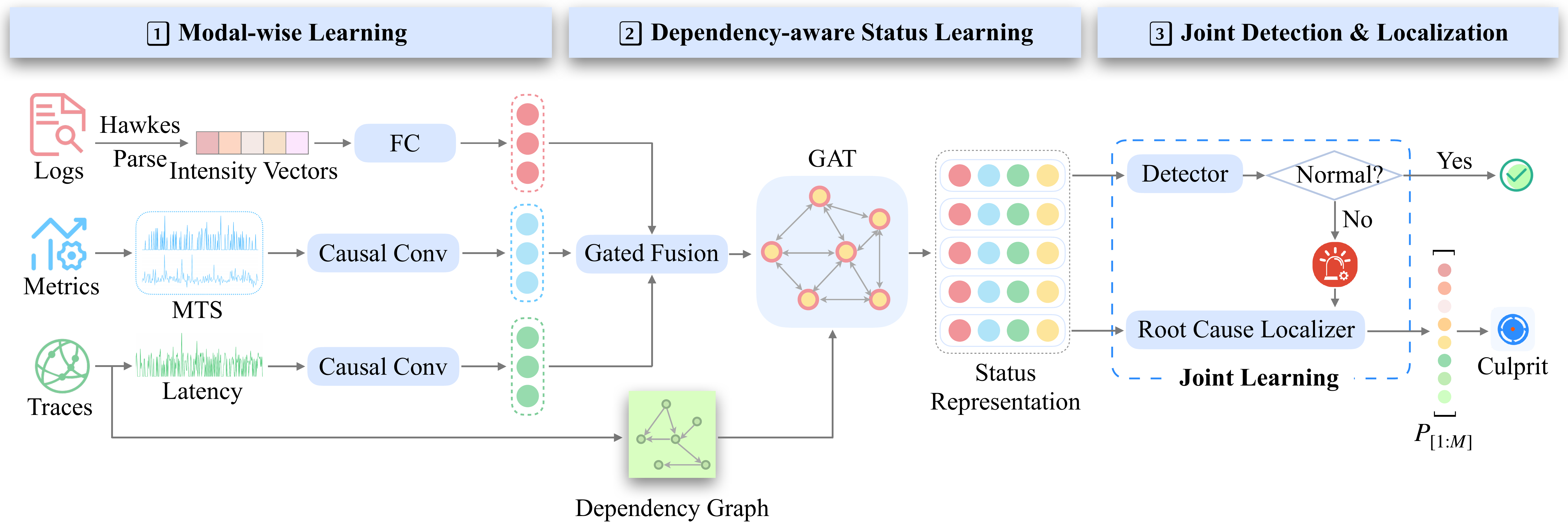}}
    \vspace{-0.2in}
    \caption{Overview of \name}
    \label{fig:overview}
 \vspace{-0.2in}
\end{figure*}

The core idea of \name is to learn the intra-service behaviors based on multi-modal data and capture dependencies between microservices to infer a comprehensive picture of the system status.
Figure~\ref{fig:overview} displays the overview of \name, containing three phases: \textit{modal-wise learning}, \textit{dependency-aware status learning}, and \textit{joint detection and localization}. 

\subsection{Modal-wise Learning}\label{sec:method:1}
This phase aims to model the different sources of monitoring data individually. We apply modality-specific models to learn an informative representation for each modality.

\subsubsection{Log Event Learning}\label{sec:method:log}
We observe that both log semantics and event occurrences can reflect anomalies (\cref{sec:study:multi-source}), yet we herein focus on event occurrences because of two reasons: 
1) the logging behavior of microservices highly relies on the developers' expertise, so the quality of log semantics cannot be guaranteed~\cite{SurveyHe};
2) the complexity of microservices necessitates lightweight techniques.
As semantic extraction requires computation-intensive natural language processing technologies, log semantic-based methods may pose challenges in practical usage.

Therefore, we focus on modeling the occurrences of log events instead of log semantics. 
An insight facilitates the model. We observe that the past event increases the likelihood of the event's occurrence in the near future, which fits the assumption of the self-exciting process~\cite{self-exciting}. 
Hence, we initially propose to adopt the Hawkes process~\cite{Hawkes}, a kind of self-exciting point process, to model the event occurrences, which is defined by the conditional intensity function:
\begin{equation}\label{eq:hawkes}
    \lambda_l^*(t) = \mu_l(t)+\sum_{\tau < t} \phi_l(t-\tau)
\end{equation}
where $l=1,...,L$ and $L$ is the number of event types; for the $l$-th event, $\mu_l$ is an estimated parameter and $\phi_l (\cdot)$ is a user-defined triggering kernel function. 
We use an exponential parametrisation of the kernels herein following~\cite{HawkesADM4}: $\phi_l (\cdot) = \alpha_l\beta \exp(-\beta t)|_{t>0}$, where $\alpha_1 \cdots \alpha_L$ are estimated parameters and $\beta$ is a hyper-parameter.

In brief, log learning is done in a three-step fashion:
\begin{itemize}[leftmargin=12pt, topsep=3pt]
\setlength\itemsep{0.4em}
\item[A.] Parsing: \name starts with parsing logs into events via Drain~\cite{Drain} by removing variables in log messages.

\item[B.] Estimating: we then record the timestamps of event occurrences (relative to the starting timestamp of the observation window) to estimate the parameters of the Hawkes model with an exponential decay kernel. The estimation is implemented via an open-source toolkit Tick~\cite{tick}.
In this way, events $\mathbf{X}^\mathcal{L}$ at each microservice inside a window are transformed into an intensity vector $\boldsymbol{\Lambda}=[\lambda_1^*,\cdots,\lambda_L^*] \in \mathbb{R}^L$. 

\item[C.] Embedding: the intensity vector $\boldsymbol{\Lambda}$ is embedded into a dense vector $\boldsymbol{H}^\mathcal{L} \in \mathbb{R}^{E^\mathcal{L}}$ in the latent space via a fully connected layer with the hidden size of ${E^\mathcal{L}}$.
\end{itemize}

\subsubsection{KPI Learning}\label{sec:method:KPI}
We first organize the KPIs $\mathbf{X}^\mathcal{K}$ with $k$ indicators of each microservice into a $k$-variate time series with the length of $T$.
Then we use a 1D dilated causal convolution (DCC)~\cite{TCN} layer that is lightweight and parallelizable to learn the temporal dependencies and cross-series relations of KPIs. 
Previous studies have demonstrated DCC's computational efficiency and accuracy in feature extraction of time series~\cite{TCNexp}.
Afterward, we apply a self-attention~\cite{Attn} operation to compute more reasonable representations, and the attention weights are as computed in Equation~\ref{eq:attn}.
\begin{equation}\label{eq:attn}
    Attn(X) = \mathsf{softmax} \left( \frac{W_qX \cdot (W_kX)^{\mathrm{T}}}{\sqrt{d}}(W_vX) \right) 
\end{equation}
where $W_q$, $W_k$, and $W_v$ are learnable parameters, and $d$ is an empirical scaling factor.
This phase outputs $\boldsymbol{H}^\mathcal{K} \in \mathbb{R}^{E^\mathcal{K}}$ representing KPIs, where $E^\mathcal{K}$ is the number of convolution filters.

\subsubsection{Trace Learning}\label{sec:method:trace}
Inspired by previous works~\cite{AID, DyCause, TraceAnomaly}, we extract latency from trace files and transform it into a time series by calculating the average latency at a time slot for each callee. We obtain a $T$-length univariate latency time series at each microservice (i.e., callee).
Similarly, the latency time series is fed into a 1D DCC layer followed by a self-attention operation to learn the latent representation $\boldsymbol{H}^\mathcal{T} \in \mathbb{R}^{E^\mathcal{T}}$, where $E^\mathcal{T}$ is the pre-defined number of filters.
Note that we simply pad time slots without corresponding invocations with zeros.

\subsection{Dependency-aware Status Learning}\label{sec:method:2}
In this phase, we aim to learn microservices' overall status and draw a comprehensive picture of the system. This module consists of three steps: dependency graph construction, multi-modal fusion, and dependency graph modeling.
We first extract a directional graph depicting the relationships among microservices from historical traces. 
Afterward, we fuse the multi-modal representations obtained from the previous phases into latent node embeddings to represent the service-level status. Messages within the constructed graph will be propagated through a graph neural network so as to learn the neighboring dependencies represented in the edge weights. 
Eventually, we can obtain a dependency-aware representation representing the overall status of the microservice system.

\subsubsection{Dependency Graph Construction}
By regarding microservices as nodes and invocations as directional edges, we can extract a dependency graph $\mathcal{G}=\{\mathbb{V}, \mathbb{E}\}$ from historical traces to depict the dependencies between microservices. 
Specifically, 
$\mathbb{V}$ is the node set and $|\mathbb{V}|=M$, where $M$ is the number of microservices; 
$\mathbb{E}$ is the set of edges, and $\vec{e}_{a,b}=(v_a, v_b) \in \mathbb{E}$ denotes an edge directed from $v_a$ to $v_b$, that is, $v_b$ has invoked $v_a$ at least once in the history.

\subsubsection{Multi-modal Fusion}\label{sec:method:fuse}
In general, there are three fusion strategies~\cite{fusion}: early fusion carried out at the input level, intermediate fusion for fusing cross-modal representations, and late fusion at the decision level (e.g., voting). 
Research in cross-modal learning ~\cite{feifei14fusion, NIPS16fusion} and neuroscience~\cite{Neuro19fusion, Multisensory21fusion} suggests that intermediate fusion usually facilitates modeling, so we transform single-modal representations to a compact multi-modal representation via intermediate fusion.

The fusion contains two steps:
\begin{itemize}[leftmargin=12pt, topsep=3pt]
\setlength\itemsep{0.4em}
\item[A.] We concatenate ($[\cdot || \cdot]$) all representations of each microservice obtained from the previous phase to retain exhaustive information. 
The resulting vector $[\boldsymbol{H}^\mathcal{L} || \boldsymbol{H}^\mathcal{K} || \boldsymbol{H}^\mathcal{T}]$ is subsequently fed into a fully connected layer to be projected into a lower-dimensional space, denoted by $\boldsymbol{H}^{\prime \mathcal{S}} \in \mathbb{R}^{2E}$, where $2E < E^\mathcal{L}+E^\mathcal{K}+E^\mathcal{T}$ is an even number.

\item[B.] $\boldsymbol{H}^{\prime \mathcal{S}}$ passes through a Gated Linear Unit (GLU)~\cite{GLU} to fuse representations in a non-linear manner and filter potential redundancy.
GLU controls the bandwidth of information flow and diminishes the vanishing gradient problem. It also possesses extraordinary resilience to catastrophic forgetting. 
As we have massive data and complex stacked neural layers, GLU fits our scenario well. 
The computation follows $\boldsymbol{H}^\mathcal{S} = GLU(\boldsymbol{H}^{\prime \mathcal{S}}) = \boldsymbol{H}^{\prime \mathcal{S}}_{(1)} \otimes \sigma(\boldsymbol{H}^{\prime \mathcal{S}}_{(2)})$, where $\boldsymbol{H}^{\prime \mathcal{S}}_{(1)}$ is the first half of $\boldsymbol{H}^{\prime \mathcal{S}}$ and $\boldsymbol{H}^{\prime \mathcal{S}}_{(2)}$ is the second half; $\otimes$ denotes element-wise product, and $\sigma$ is a sigmoid function. 
\end{itemize}
Finally, we obtain $\boldsymbol{H}^\mathcal{S} \in \mathbb{R}^{E}$, a service-level representation of each microservice.

\subsubsection{Dependency Graph Learning}\label{sec:method:graph}
As interactions between microservices can be naturally described by dependency graphs, we apply graph neural networks to perform triage inference.
Particularly, we employ Graph Attention Network (GAT)~\cite{GAT-v2} to learn the dependency-aware status of the microservice system.
GAT enables learning node and edge representations and dynamically assigns weights to neighbors without requiring computation-consuming spectral decompositions.
Hence, the model can pay attention to microservices with abnormal behaviors or at the communication hub.

The local representation $\boldsymbol{H}^\mathcal{S}$ serves as the node feature, and GAT learns the whole graph's representation, where dynamic weights of edges are computed as Equation~\ref{eq:gat}.
\begin{equation}\label{eq:gat}
    \omega_{a,b} = \frac{\exp(\mathsf{LeakyReLU}
    (v^\mathrm{T}
    [\boldsymbol{W}\boldsymbol{H}_a^\mathcal{S} || \boldsymbol{W}\boldsymbol{H}_b^\mathcal{S}]))}
    {\sum_{k \in \mathbb{N}_a} 
    \exp(\mathsf{LeakyReLU}(v^\mathrm{T}
    [\boldsymbol{W}\boldsymbol{H}_a^\mathcal{S} || \boldsymbol{W}\boldsymbol{H}_k^\mathcal{S}]))}
\end{equation}
where $\omega_{a,b}$ is the computed weight of edge $\vec{e}_{a,b}$;
$\mathbb{N}_a$ is the set of neighbor nodes of node $a$; $\boldsymbol{H}_a^\mathcal{S}$ is the inputted node feature of $a$; 
$\boldsymbol{W} \in \mathbb{R}^{E^\mathcal{G} \times E}$ and $v \in \mathbb{R}^{2E^\mathcal{G}}$ are learnable parameters.
$E^\mathcal{G}$ is the dimension of the outputted representation, which is calculated by $\boldsymbol{\hat{H}}_a^{\mathcal{S}} = \psi (\sum_{b \in \mathbb{N}_a} \omega_{a,b} \boldsymbol{W}\boldsymbol{H}_b^{\mathcal{S}})$, where $\psi(\cdot)$ is a customized activation function, usually ReLU.
Eventually, we perform global attention pooling~\cite{GAP} on the multi-modal representations of all nodes. The final output is $\boldsymbol{H}^\mathcal{F} \in \mathbb{R}^{E^\mathcal{F}}$, a dependency-aware representation of the overall system status.

\subsection{Joint Detection and Localization}\label{sec:method:output}
Lastly, \name predicts whether the current observation window is abnormal and if so, it identifies which microservice the root cause is.
As demonstrated in~\cref{sec:study:detection}, existing troubleshooting methods regard anomaly detection and root cause localization as independent and ignore their shared knowledge. 
Besides, current anomaly detectors deliver unsatisfactory results and affect the next-stage localization by incorporating noisy labels.
Therefore, we fully leverage the shared knowledge and integrate two closely related tasks into an end-to-end model.
\begin{figure}[htb]
    \centering
    \vspace{-0.1in}
        {\includegraphics[width=0.99\linewidth]{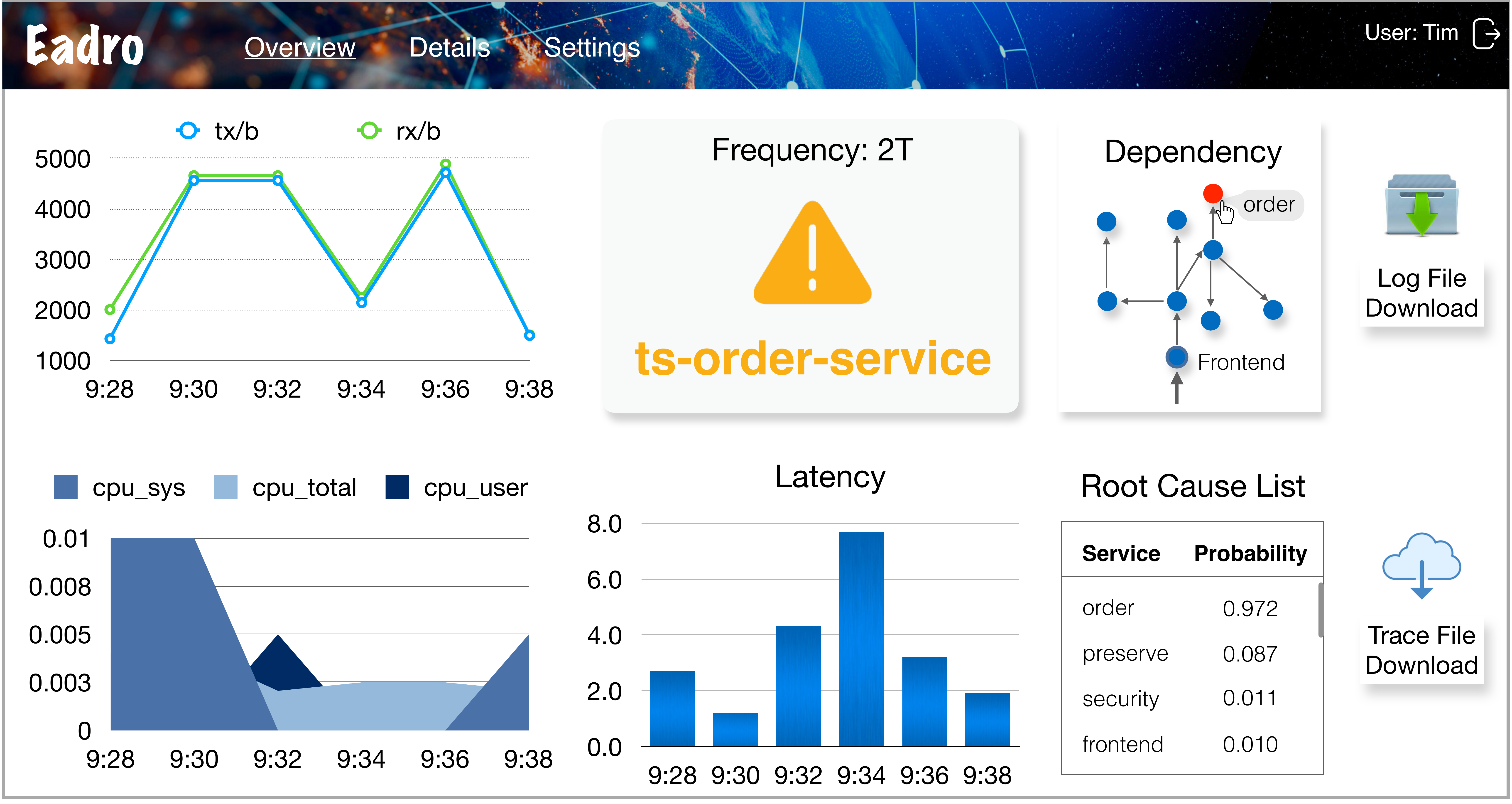}}
    \vspace{-0.2in}
    \caption{A demo for reviewing the suspicious status.}
    \label{fig:interface}
\vspace{-0.1in}
\end{figure}

In particular, based on the previously obtained representation $\boldsymbol{H}^\mathcal{F}$, a detector first conducts binary classification to decide the existence of anomalies.
If no anomaly exists, \name directly outputs the result; if not, a localizer ranks the microservices according to their probabilities of being the culprit.
The detector and the localizer are both composed of stacked fully-connected layers and jointly trained by sharing an objective.
The detector aims to minimize the binary cross-entropy loss:
\begin{equation}\label{eq:loss-detection}
    \mathfrak{L}_1 = \sum_{i=1}^N [-(y_i\log(\hat{y_i}) + (1-y_i)\log(1-\hat{y_i}))]
\end{equation}
where $N$ is the number of historical samples; $y_i \in \{0,1\}$ is the ground truth indicating the presence of anomalies (1 denotes presence while 0 denotes absence), and $\hat{y_i} \in [0,1]$ is the predicted indicator.
Subsequently, all samples predicted as normal (0) are masked, and samples predicted as abnormal (1) pass through the localizer. 
The localizer attempts to narrow the distance between the predicted and ground-truth probabilities, whose objective is expressed by:
\begin{equation}\label{eq:loss-localization}
    \mathfrak{L}_2 = \sum_{i=1}^N \sum_{s=1}^M c_{i, s} \log (p_{i,s})
\end{equation}
where $M$ is the number of involved microservices. In the $i$-th sample, $c_{i,s} \in \{0,1\}$ is 1 if the culprit microservice is $s$ and 0 otherwise; $p_{i,s}$ is the predicted probability of microservice $s$ being the culprit.
The objective of \name is the weighted sum of the two sub-objectives $\mathfrak{L} = \beta \cdot \mathfrak{L}_1 + (1-\beta) \cdot \mathfrak{L}_2$, where $\beta$ is a hyper-parameter balancing the two tasks.
Eventually, \name outputs a ranked list of microservices to be checked according to their predicted probabilities of being the root cause.

To sum up, \name can provide explicit clues about the microservice status. Hence, troubleshooting is much more convenient for operation engineers with the ranked list of microservices.
Figure~\ref{fig:interface} presents a visualized demo.

\section{Evaluation}
This section answers the following research questions:
\begin{itemize}[leftmargin=12pt, topsep=0pt]
    \item \textbf{RQ1}: How effective is \name in anomaly detection?
    \item \textbf{RQ2}: How effective is \name in root cause localization?
    \item \textbf{RQ3}: How much does each data source contribute?
\end{itemize}

\subsection{Data Collection}\label{sec:data}
Since existing data collections of microservice systems~\cite{AiopsChallenge20, TraceData20} contain traces only, we deploy two benchmark microservice systems and generate requests to collect multi-source data, including logs, KPIs, and traces.
Afterward, we inject typical faults to simulate real-world anomalies.
To our best knowledge, it is the first triple-source data collection 
 with injected faults in the context of microservices.

\subsubsection{Benchmark microservice systems}
We first deploy two open-source microservice benchmarks: TrainTicket~\cite{TrainTicket} (TT) and SocialNetwork~\cite{SocialNetwork} (SN).
TT provides a railway ticketing service where users can check, book, and pay for train tickets. It is widely used in previous works~\cite{MEPFL, TraceAnomaly} with 41 microservices actively interacting with each other, and 27 of them are business-related.
SN implements a broadcast-style social networking site. Users can create, read, favorite, and repost posts. In this system, 21 microservices communicate with each other via Thrift RPCs~\cite{Thrift}.
SN has 21 microservices, 14 of which are related to business logic. 
 
We construct a distributed testbed to deploy the two systems running in Docker containers and develop two request simulators to simulate valid user requests.
A series of open-source monitoring tools are deployed for data collection.
Microservice instances send causally-related traces to a collector Jaeger~\cite{Jaeger}.
We employ cAdvisor~\cite{cAdvisor} and Prometheus~\cite{Prometheus} to monitor the KPIs per second of each microservice.
The KPIs are stored in an instance of InfluxDB~\cite{Influxdb}, including ``CPU system usage'', ``CPU total usage'', ``CPU user usage'', ``memory usage'', the amount of ``working set memory'', ``rx bytes'' (received bytes), and ``tx bytes'' (transmitted bytes). 
We also utilize Elasticsearch~\cite{Elasticsearch}, Fluentd~\cite{Fluentd}, and Kibana~\cite{Kibana} to collect, aggregate, and store logs, respectively.

\subsubsection{Fault Injection}
\name can troubleshoot anomalies that manifest themselves in performance degradations (logs and KPIs) or latency deviations (traces). Referring to previous studies~\cite{MicroRank, AutoMAP, DyCause}, we inject three typical types of faults via Chaosblade~\cite{Chaosblade}. 
Specifically, we simulate CPU exhaustion by putting a hog to consume CPU resource heavily. 
To simulate a network jam, we delay the network packets of a microservice instance. 
We also randomly drop network packets to simulate stochastic packet loss that frequently occurs when excessive data packets flood a network.

We generate 0.2$\sim$0.5 and 2$\sim$3 requests per second for TT and SN at a uniform rate, respectively.
Before fault injection, we collect normal data under a fault-free setting for 7 hours for TT and 1.2 hours for SN.
Then, we set each fault duration to 10 mins (with a 2-min interval between two injections) for TT, while the fault duration is 2 mins and SN's interval is half a minute. 
Each fault is injected into one microservice once.
In total, we conduct 162 and 72 injection operations in TT and SN, respectively.
Such different setups are attributed to the different processing capacities of the two systems, i.e., TT usually takes more time to process a request than SN.

In this way, we collect two datasets (\A and \B) with 48,296 and 126,384 traces, respectively. Data produced in different periods are divided into training (60\%) data and testing (40\%) data, respectively. The data divisions share similar distributions in abnormal/normal ratios and root causes.

\subsection{Baselines} 
We compare \name with previous approaches and derived methods integrating multi-source data. As our task is relatively novel by incorporating more information than existing single-source data-based studies, simply comparing our model with previous approaches seems a bit unfair.

\subsubsection{Advanced baselines} 
In terms of anomaly detection, we consider two state-of-the-art baselines.
TraceAnomaly~\cite{TraceAnomaly} uses a variational auto-encoder (VAE) to discover abnormal invocations. 
MultimodalTrace~\cite{MMTrace} extracts operation sequences and latency time series from traces and uses a multi-modal Long Short-term Memory (LSTM) network to model the temporal features.
For root cause localization, we compare \name with five trace-based baselines: TBAC~\cite{TBAC}, NetMedic~\cite{NetMedic}, MonitorRank~\cite{MonitorRank}, CloudRanger~\cite{CloudRanger}, and DyCause~\cite{DyCause}.
As far as we know, no root cause localizers for microservices rely on multi-modal data. 

These methods use statistical models or heuristic methods to locate the root cause. For example, TBAC, MonitorRank, and DyCause applied the Pearson correlation coefficient, and MonitorRank and DyCause also leveraged Random Walk.
We implement these baselines referring to the codes provided by the original papers~\cite{TraceAnomaly, DyCause, MicroRank}. For the papers without open-source codes, we carefully follow the papers and refer to the baseline implementation released by~\cite{DyCause}.

\subsubsection{Derived multi-source baselines} \label{sec:setup:derived}
We also derive four multi-source data-based methods for further comparison.
Inspired by~\cite{MMTrace}, we transform all data sources into time series and use learning-based algorithms for status inference. 
Specifically, logs are represented by event occurrence sequences; traces are denoted by latency time series; KPIs are natural time series.
Since previous studies are mainly machine learning-based, we train practical machine learning methods, i.e., Random Forest (RF) and Support Vector Machine (SVM), on the multi-source time series.
We derive MS-RF-AD and MS-SVM-AD for anomaly detection as well as MS-RF-RCL and MS-SVM-RCL for root cause localization.
We also derive two methods (MS-LSTM and MS-DCC) that employ deep learning techniques, i.e., LSTM and 1D DCC, to extract representations from multi-modal time series.
The learned representations are fed into the module of joint detection and localization, which is described in~\ref{sec:method:output}.

\subsection{Implementation}
The experiments are conducted on a Linux server with an NVIDIA GeForce GTX 1080 GPU via Python 3.7. 
As for the hyper-parameters, the hidden size of all fully-connected layers is 64, and every DCC layer shares the same filter number of 64 with a kernel size of three. The GAT's hidden size and the fusion dimension (i.e., $2E$) are 128.
We use a 4-head mechanism of GAT's attention layer, and the layer number of all modalities' models is only one for speeding up.
Moreover, Batch Normalization~\cite{BatchNorm} is added after DCCs to mitigate overfitting.
We train \name using the Adam~\cite{Adam} optimizer with an initial learning rate of 0.001, a batch size of 256, and an epoch number of 50.
All the collected data and our code are released for replication.

\subsection{Evaluation Measurements}
The anomaly detection challenge is modeled in a binary classification manner, so we apply the widely-used binary classification measurements to gauge the performance of models:
Recall (\textit{Rec})$=\frac{TP}{TP+FN}$, Precision (\textit{Pre})$=\frac{TP}{TP+FP}$, F1-score (\textit{F1})$=\frac{2 \cdot Pre \cdot Rec}{Pre+Rec}$,
where $TP$ is the number of discovered abnormal samples; $FN$ and $FP$ are defined in~\cref{sec:study:detection}.

For root cause localization, we introduce the Hit Rate of top-k (\textit{HR@k}) and Normalized Discounted Cumulative Gain of top-k (\textit{NDCG@k}) for localizer evaluation. Herein, we set $k=1,3,5$.
\textit{HR@k}$=\frac{1}{N} \sum_{i=1}^N(s_i^t \in S^{p}_{i, [1:k]})$ calculates the overall probability of the culprit microservice within the top-k predicted candidates $S^{p}_{i, [1:k]}$, where $s_i^t$ is the ground-truth root cause for the $i$-th observation window, and $N$ is the number of samples to be tested. 
\textit{NDCG@k}$=\frac{1}{N} \sum_{i=1}^N (\sum_{j=1}^M \frac{p_j}{\log_2(j+1)})$ measures the ranking quality, where $p_j$ is the predicted probability of the $j$-th microservice, and $M$ is the number of microservices. 
\textit{NDCG@1} is left out because it is the same with \textit{HR@1} in our scenario.
The two evaluation metrics measure how easily engineers find the culprit microservice. \textit{HR@k} directly measures how likely the root cause will be found within $k$ checks. \textit{NDCG@k} measures to what extent the root cause appears higher up in the ranked candidate list.
Thus, the higher the above measurements, the better. 

\subsection{\textbf{RQ1}: Effectiveness in Anomaly Detection}
\begin{table}[htbp]
\small
\centering
\vspace{-0.15in}
\caption{Performance Comparison for Anomaly Detection}
\begin{adjustbox}{max width=\columnwidth}
\begin{tabular}{l*{7}{c}}
\toprule
\multirow{2}*{\centering \textbf{Approaches}} & \multicolumn{3}{c}{\A} & \multicolumn{3}{c}{\B}\\
\cmidrule(lr){2-4}\cmidrule(lr){5-8}
&\textit{F1}&\textit{Rec}&\textit{Pre}&\textit{F1}&\textit{Rec} &\textit{Pre}\\
\cmidrule{1-8}\morecmidrules\cmidrule{1-8}
TraceAnomaly & 0.486 & 0.414 & 0.589 & 0.539 & 0.468 & 0.636 \\
MultimodalTrace & 0.608 & 0.576 & 0.644 & 0.676 & 0.632 & 0.726\\
\hdashline[2pt/5pt]
MS-RF-AD & 0.817 & 0.705 & 0.971 & 0.773 & 0.866 & 0.700 \\
MS-SVM-AD & 0.787 & 0.678 & 0.938 & 0.789 & 0.770 & 0.808\\
MS-LSTM & 0.967 & 0.997 & 0.940 & 0.948 & 0.959 & 0.937 \\
MS-DCC & 0.965 & 0.993 & 0.938  & 0.948 & 0.962 & 0.934 \\
\midrule
\textbf{\name} & \textbf{0.989} & \textbf{0.995} & \textbf{0.984} & \textbf{0.986} & \textbf{0.996} & \textbf{0.977}\\
%F1: 0.9875, 
\bottomrule
\end{tabular}
\end{adjustbox}
\label{tab:exp:detection}
\end{table}

Ground truths are based on the known injection operations, i.e., if a fault is injected, then the current observation window is abnormal; otherwise, it is normal. Table~\ref{tab:exp:detection} displays a comparison of anomaly detection, from which we draw three observations:

(1) \name outperforms all competitors significantly and achieves very high scores in \textit{F1} (0.988), \textit{Rec} (0.996), and \textit{Pre} (0.981), illustrating that \name generates very few missing anomalies or false alarms.
\name's excellence can be attributed to 1) \name applies modality-specific designs to model various sources of data as well as a multi-modal fusion to wrangle these modalities so that it can learn a distinguishable representation of the status; 2) \name learns dependencies between microservices to enable extraction of anomaly propagation to facilitate tracing back to the root cause.

(2) Generally, multi-source data-based approaches, including \name, perform much better than trace-relied baselines because they incorporate extra essential information (i.e., logs and KPIs) besides traces. 
The results align with our observations in~\cref{sec:study:multi-source} that logs and KPIs provide valuable clues about microservice anomalies, while traces cannot reveal all anomalies.
Trace-based methods can only detect anomalies yielding an enormous impact on invocations, so they ignore anomalies reflected by other data sources.

(3) Moreover, \name, MS-LSTM, and MS-DDC perform better than MS-SVM and MS-RF. 
The superiority of the former ones lies in applying deep learning and joint learning. 
Deep learning has demonstrated a powerful capacity in extracting features from complicated time series~\cite{TCNexp, DeepTSCreview, TSDMreview}. 
Joint learning allows capturing correlated knowledge across detection and localization to exploit commonalities across the two tasks.
These two mechanisms are beneficial to troubleshooting by enhancing representation learning.

In brief, \name is very effective in anomaly detection of microservice systems and improves \textit{F1} by 53.82\%$\sim$92.68\% compared to baselines and 3.13\%$\sim$25.32\% compared to derived methods. 
The detector is of tremendous assistance for next-stage root cause localization by reducing noisy labels inside the localizer's inputs.

\subsection{\textbf{RQ2}: Effectiveness in Root Cause Localization}
\begin{table*}[htbp]
\small
\centering
\caption{Performance Comparison for Root Cause Localization}
\begin{tabular}{l*{10}{c}}
\toprule
\multirow{2}*{\centering \textbf{Approaches}} & \multicolumn{5}{c}{\A} & \multicolumn{5}{c}{\B}\\
\cmidrule(lr){2-6}\cmidrule(lr){7-11}
&\textit{HR@1}&\textit{HR@3}&\textit{HR@5}&\textit{NDCG@3}&\textit{NDCG@5}&\textit{HR@1}&\textit{HR@3}&\textit{HR@5}&\textit{NDCG@3}&\textit{NDCG@5} \\
\cmidrule{1-11}\morecmidrules\cmidrule{1-11}
TBAC & 0.037 & 0.111 & 0.185 & 0.079 & 0.109 & 0.001 & 0.085 & 0.181 & 0.048 & 0.087 \\
NetMedic & 0.094 & 0.257 & 0.425 & 0.195 & 0.209 & 0.069 & 0.187 & 0.373 & 0.146 & 0.218 \\
MonitorRank & 0.086 & 0.199 & 0.331 & 0.142 & 0.196 & 0.068 & 0.118 & 0.221 & 0.095 & 0.137\\
CloudRanger & 0.101 & 0.306 & 0.509 & 0.218 & 0.301 & 0.122 & 0.382 & 0.629 & 0.269 & 0.370\\
DyCause & 0.231 & 0.615 & 0.808 & 0.448 & 0.607 & 0.273 & 0.636 & 0.727 & 0.301 & 0.353 \\
\hdashline[1pt/5pt]
MS-RF-RCL & 0.637 & 0.922 & \underline{0.970} & 0.807 & 0.827 &0.704 & \underline{0.908} & \underline{0.970} & 0.825 & \underline{0.851}\\
MS-SVM-RCL& 0.541 & 0.908 & 0.944 & 0.814 & 0.820 & 0.614 & 0.838 & \underline{0.955} & 0.741 & 0.790 \\
MS-LSTM & 0.756 & 0.930 & \underline{0.969}	& 0.859 & 0.877 & 0.757 & \underline{0.884} & \underline{0.907} & 0.834 & \underline{0.844}\\
MS-DCC & 0.767 & 0.938 & 0.972 & 0.870 & 0.882 & 0.789 & 0.968 & 0.985 & 0.898 & 0.905\\
\midrule 
\textbf{\name} & \textbf{0.990} & \textbf{0.992} & \textbf{0.993} & \textbf{0.994} & \textbf{0.994} & \textbf{0.974} & \textbf{0.988} & \textbf{0.991} & \textbf{0.982} & \textbf{0.983}\\
\bottomrule
\end{tabular}
\vspace{-0.1in}
\label{tab:exp:localization}
\end{table*}

To focus on comparing the effectiveness of root cause localization, we provide ground truths of anomaly existence for baselines herein.
In contrast, \name, MS-LSTM, and MS-DCC use the predicted results of their detectors as they are end-to-end approaches integrating the two tasks.
Table~\ref{tab:exp:localization} presents the root cause localization comparison, underpinning three observations:

(1) \name performs the best, taking all measurements into consideration, achieving \textit{HR@1} of 0.982, \textit{HR@5} of 0.990, and \textit{NDCG@5} of 0.989 on average.
With the incorporation of valuable logs and KPIs ignored by previous approaches, \name can depict the system status more accurately.
Trace-based approaches have difficulties in troubleshooting resource exhaustion-related anomalies or severe network-related anomalies that block inter-service communications resulting in few invocations.
Besides, \name enables eavesdropping across detection and localization via joint learning, which encourages full use of the shared knowledge to enhance status learning. 
\name also leverages powerful techniques to capture meaningful patterns from multi-modal data, including designs of modality-specific models and advanced GAT to exploit graph-structure dependencies. 
Moreover, \name achieves a much higher score in \textit{HR@1} than derived methods, while its superiority in \textit{HR@5} and \textit{NDCG@5} is not particularly prominent.
The reason is that \name learns the dependency-aware status besides intra-service behaviors, allowing to catch the anomaly origin by tracing anomaly propagation.
Other multi-modal approaches capture dependency-agnostic information, so they can pinpoint the scope of suspicious microservices effectively rather than directly deciding the culprit.

(2) Multi-modal approaches considerably outperform single-modal baselines, similar to the results in anomaly detection.
The superiority of multi-source derived methods is more evident since localization is a more complicated task than detection, so the advantage of incorporating diverse data sources to learn the complementarity is fully demonstrated.
This situation is more revealing in \A because TrainTicket responds more slowly, leading to sparse trace records, and trace-based models get into trouble when few invocations occur in the current observation window.
In contrast, derived approaches can accurately locate the culprit microservice in such a system since they leverage various information sources to obtain more clues.

(3) Considering multi-modal approaches, \name, MS-LSTM, and MS-DCC deliver better performance (measured by \textit{HR@1}) than MS-RF-RCL and MS-SVM-RCL.
The superiority of the former approaches can be attributed to the strong fitting ability of deep learning and the advantages brought by the joint learning mechanism.
However, MS-LSTM performs poorer in narrowing the suspicious scope, especially in \B (measured by \textit{HR@5} and \textit{NDCG@5}).
This may be because that LSTMs' training process is a lot more complicated than DCCs or simple machine learning techniques. The scale of \B is relatively small, so MS-LSTM cannot be thoroughly trained and capture the most meaningful features.

To sum up, the results demonstrate the effectiveness of \name in root cause localization. \name increases \textit{HR@1} by 290\%$\sim$5068\% than baselines and 26.93\%$\sim$66.16\% than derived methods.
Our approach shows effectiveness both in anomaly detection and root cause localization, suggesting its potential to automate labor-intensive troubleshooting.

\subsection{\textbf{RQ3}: Contributions of Different Data Sources}
We perform an ablation study to explore how different data sources contribute by conducting source-wise-agnostic experiments, so we derive the following variants:

\begin{itemize}[leftmargin=12pt, topsep=0pt]
    \item \name w/o $\mathcal{L}$: drops logs while inputs traces and KPIs by removing the log modeling module in \cref{sec:method:log}.
    \item \name w/o $\mathcal{M}$: drops KPIs while inputs traces and logs by removing the KPI modeling module in \cref{sec:method:KPI}.
    \item \name w/o $\mathcal{T}$: drops latency extracted from traces by removing the trace modeling module in \cref{sec:method:trace}.
    \item \name w/o $\mathcal{G}$: replaces GAT by an FC layer to learn dependency-agnostic representations.
\end{itemize}

\begin{table}[htb]
\small
\centering
\vspace{-0.1in}
\caption{Experimental Results of the Ablation Study}
\begin{adjustbox}{max width=\columnwidth}
\begin{tabular}{lcccccc}
\toprule
\multirow{2}*{\centering \textbf{Variants}} & \multicolumn{3}{c}{\A} & \multicolumn{3}{c}{\B}\\
\cmidrule(lr){2-4}\cmidrule(lr){5-7}
&\textit{HR@1}&\textit{HR@5}&\textit{F1}&\textit{HR@1}&\textit{HR@5}&\textit{F1}\\
\cmidrule{1-7}\morecmidrules\cmidrule{1-7}
\textbf{\name} & \textbf{0.990} & \textbf{0.993} & \textbf{0.989} & \textbf{0.974} & \textbf{0.991}& \textbf{0.986}\\
\midrule
{\name w/o $\mathcal{L}$} & 0.926 & 0.993 & 0.964 & 0.902 & 0.954 & 0.972\\
{\name w/o $\mathcal{M}$} & 0.776 & 0.962 & 0.960 & 0.684 & 0.947 & 0.974\\
{\name w/o $\mathcal{T}$} & 0.785 & 0.930 & 0.945 & 0.627 & 0.930 & 0.957\\
{\name w/o $\mathcal{G}$} & 0.803 & 0.982 & 0.970 & 0.791 & 0.960 & 0.946 \\
\bottomrule
\end{tabular}
\end{adjustbox}
\vspace{-0.1in}
\label{tab:exp:ablation}
\end{table}

The ablation study results are shown in Table~\ref{tab:exp:ablation}. 
Considering that root cause localization, being our major target, is more difficult and that all variants achieve relatively good performance in anomaly detection, we focus on root cause localization.
Clearly, each source of information contributes to the effectiveness of \name as it performs the best, while the degrees of their contributions are not exactly the same.

Specifically, logs contribute the least as {\name~w/o~$\mathcal{L}$} is second-best.
We attribute it to the lack of log semantics and the low logging frequency. As the two benchmark systems were recently proposed without multiple version iterations, only a few events are recorded. We believe that logs would play a greater value in the development of microservices.

In addition, we observe that the performance of {\name~w/o~$\mathcal{M}$} and {\name~w/o~$\mathcal{T}$} degrades dramatically, especially in \textit{HR@1}, since traces and KPIs are essential information that contributes the most to the identification of the root cause microservice.
This observation aligns with our motivating cases, where we show some anomaly cases that can be directly revealed by traces and KPIs.

Moreover, \textit{HR@5} of {\name~w/o~$\mathcal{G}$} degrades slightly, indicating that dependency-agnostic representations are useful to narrow the suspicious scope.
However, \textit{HR@1} of {\name~w/o~$\mathcal{G}$} decreases 23.21\% as \name uses readily applicable GAT to modal graph-structure inter-service dependencies, while FC layers model the dependencies linearly, unable to capture anomaly propagation well, leading to performance degradation in determining the culprit.

\begin{figure}[htbp]
\centering
\subfigure[\name]{
\begin{minipage}{0.49\textwidth}
\centering 
\includegraphics[width=0.95\linewidth]{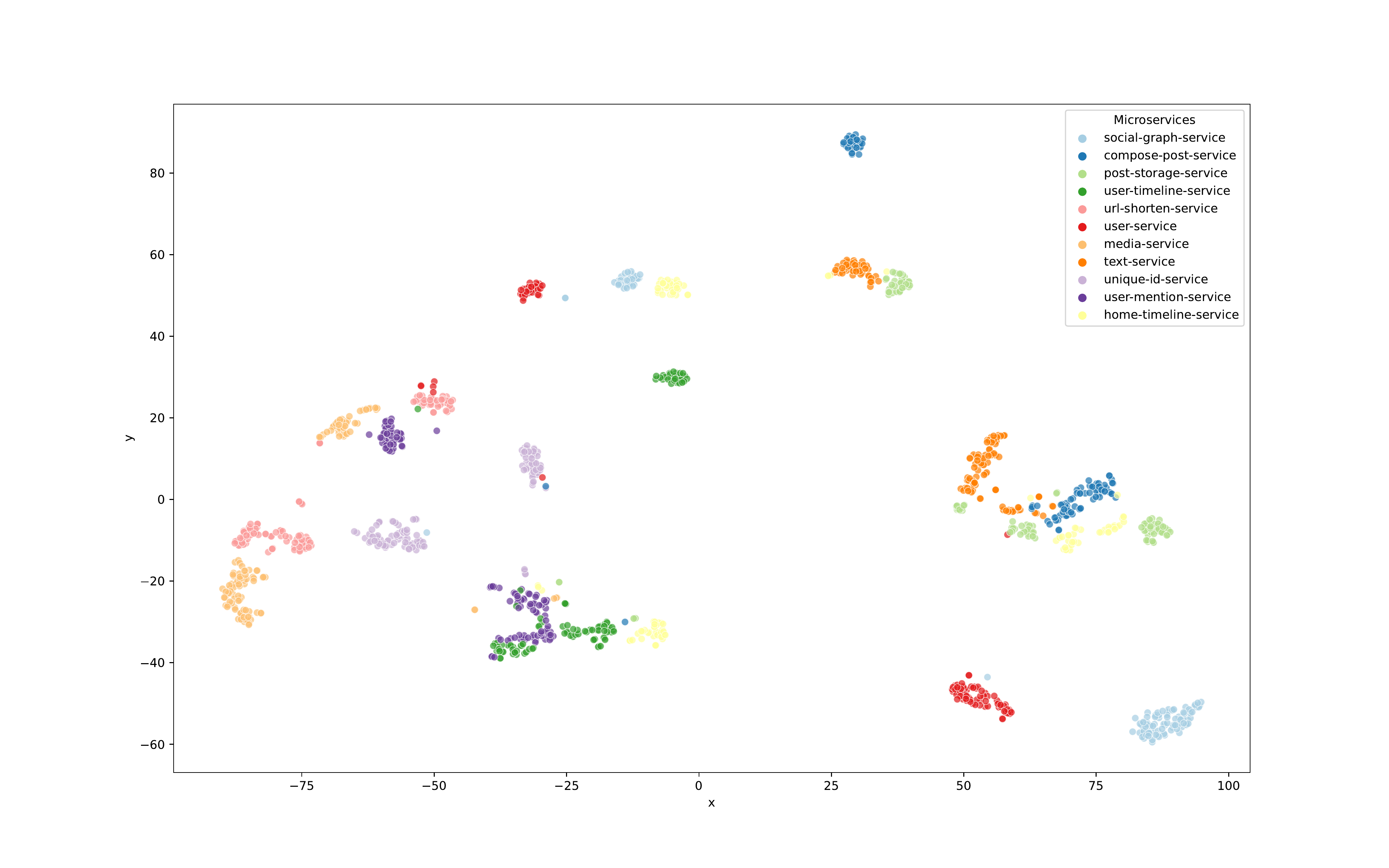}
\end{minipage}
}
\subfigure[{\name w/o $\mathcal{L}$}]{
\begin{minipage}{0.49\textwidth}
\centering 
\includegraphics[width=0.95\linewidth]{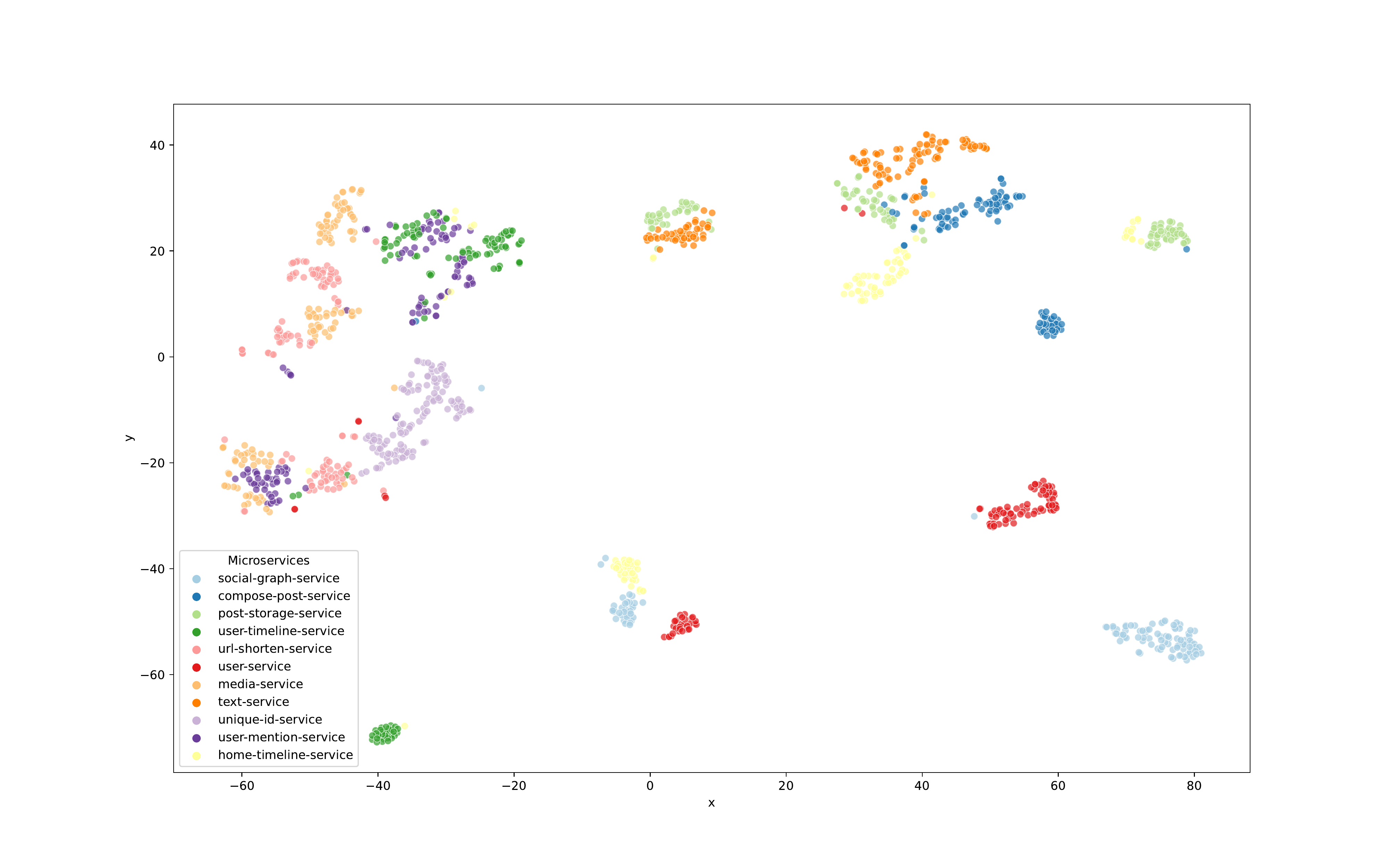}
\end{minipage}
}
\subfigure[{\name w/o $\mathcal{M}$}]{ 
\begin{minipage}{0.49\textwidth}
\centering
\includegraphics[width=0.95\linewidth]{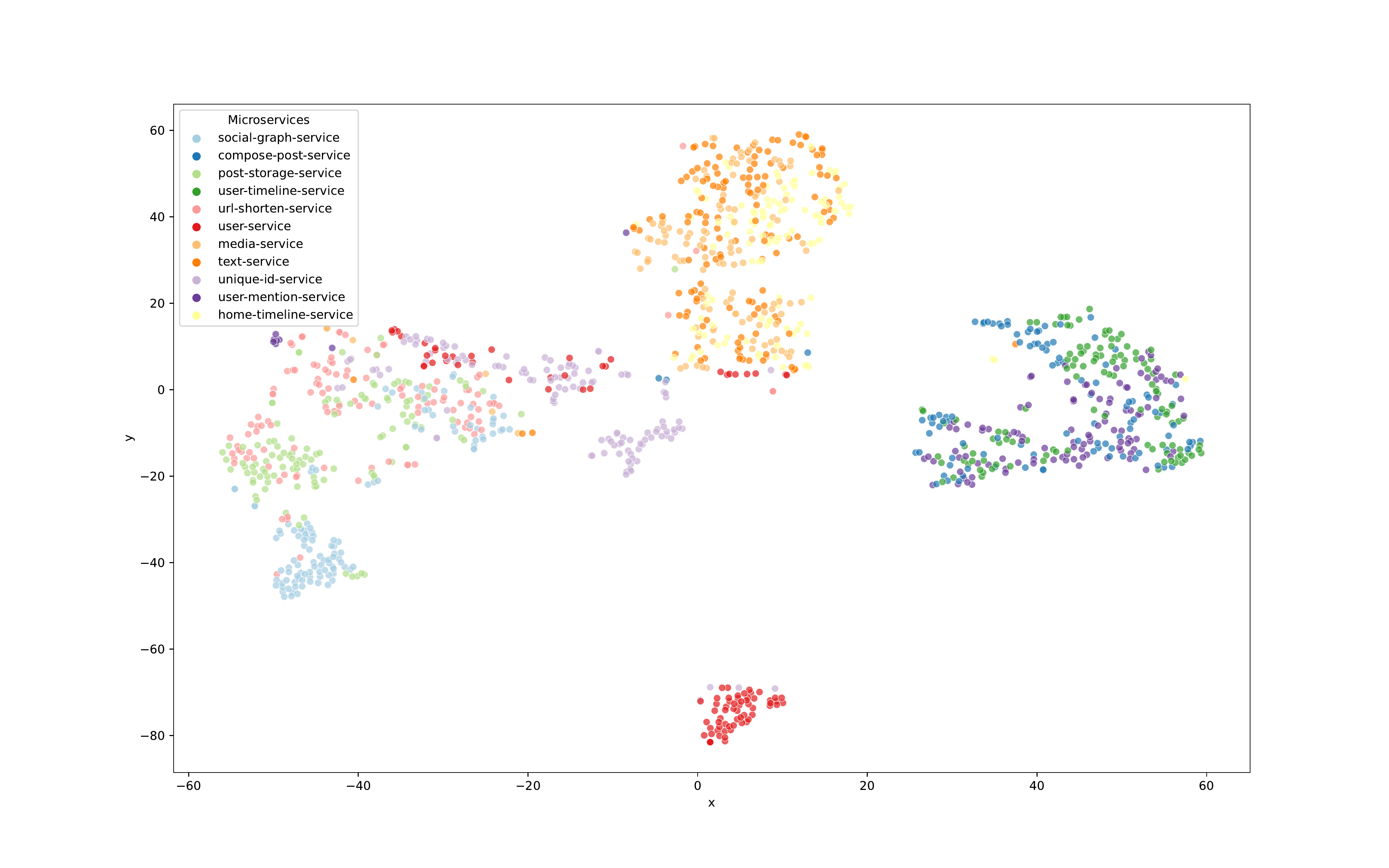}
\end{minipage}
}
\vspace{-0.1in}
\caption{Distributions of representations learned by \name and its variants.}
\vspace{-0.3in}
\label{fig:t-sne}
\end{figure}

To further demonstrate the benefits brought by KPIs and logs, we visualize the latent representations of abnormal data samples learned by \name, {\name~w/o~$\mathcal{L}$}, and {\name~w/o~$\mathcal{M}$} via t-SNE~\cite{t-SNE} of the test set of \B, shown in Figures~\ref{fig:t-sne}.

We can see that the representations learned by \name are the most discriminative, and those learned by {\name~w/o~$\mathcal{L}$} are second-best, while those learned by {\name~w/o~$\mathcal{M}$} are the worst. 
Specifically, \name distributes representations corresponding to the different root causes into different clusters distant from each other in the hyperspace.
In contrast, {\name~w/o~$\mathcal{M}$} learns representations close in space, making it difficult to distinguish them for triage. That is why {\name~w/o~$\mathcal{M}$} delivers poorer performance in localization than \name. The visualization intuitively helps us grasp the usefulness of KPIs in helping pinpoint the root cause.
The discriminativeness of the representations learned by {\name~w/o~$\mathcal{L}$} is in-between, where some clusters are pure while others seem to be a mixture of representations corresponding to different root causes, in line with the experiment results.
We can attribute part of the success of \name to incorporating KPIs and logs, which encourages more discriminative representations of the microservice status with extra clues.

In conclusion, the involved data sources can all contribute to the effectiveness of \name to some degree, and traces contribute the most to the overall effectiveness. This emphasizes the insights about appropriately modeling multi-source data to troubleshoot microservices effectively.

\section{Disucussion}\label{sec:threats}
\subsection{Limitations}
We identify three limitations of \name: 1) the incapacity to deal with bugs related to program logic; 2) the prerequisites for multi-source data collection; 3) the requirement of annotated data for training.

As \name is an entirely data-driven approach targeting the scope of reliability management, it is only applicable to troubleshooting anomalies manifested in the involved data, so logical bugs out of our scope and silent issues that do not incur abnormal patterns in observed data can not be detected or located.

Moreover, \name is basically well-suited for all microservices where anomalies can be reflected in the involved three types of data we employ. However, some practical systems may lack the ability to collect the three types of data. Though the low-coupled nature of the modal-wise learning module allows the absence of some source of data, it is better to provide all data types to fully leverage \name.
Since we apply standard open-source monitoring toolkits and these off-the-shelf toolkits can be directly instrumented, enabling microservices with the data collection ability is not difficult.

In addition, the supervised nature of \name requires a large amount of labeled training data, which may be time-consuming in the real world. 
Nevertheless, our approach outperforms compared with unsupervised approaches by a large margin, indicating that in practice, unsupervised methods may be difficult to use because the accuracy rate is not up to the required level, especially considering that realistic microservices systems are much larger and more complex.
A common solution in companies is to use an unsupervised model to generate coarse-grained pseudo-labels. Afterward, experienced engineers manually review the labels with lower confidence.
The hybrid-generated labels are used for training the supervised model, and eventually, the supervised approach performs the troubleshooting work.
Hence, \name will still play an important role in practice and fulfill its potential.

\subsection{Threat to Validity}
\subsubsection{Internal Threat}
The main internal threat lies in the correctness of baseline implementation.
We reproduce the baselines based on our understanding of their papers since most baselines, except DyCause and TraceAnomaly, have not released codes, but the understanding may not be accurate.
To mitigate the threat, we carefully follow the original papers and refer to the baseline implementation released by~\cite{DyCause}. 

\subsubsection{External Threat}
The external threats concern the generalizability of our experimental results. We evaluate our approach on two simulated datasets since there is no publicly available dataset containing multi-modal data.
It is yet unknown whether the performance of \name can be generalized across other datasets.
We alleviate this threat from two aspects.
First, the benchmark microservice systems are widely used in existing comparable studies, and the injected faults are also typical and broadly applied in previous studies~\cite{MicroRank, AutoMAP, DyCause}, thereby supporting the representativeness of the datasets.
Second, our approach is request- and fault-agnostic, so an anomaly incurred by a fault beyond our injections can also be discovered if it causes abnormalities in the observations.

\section{Related Work}\label{sec:relatedwork}
Previous anomaly detection approaches are usually based on system logs~\cite{Deeplog, LogAnomaly, LogRobust, SwissLog, BertLog} or KPIs~\cite{Adsketch, SRCNN, InterFusion, USAD, OmniAnomaly}, or both~\cite{Hades}, targeting traditional distributed systems without complex invocation relationships.
Recently, some studies~\cite{TraceAnomaly, MMTrace, AID} have been presented to automate anomaly detection in microservice systems. 
\cite{TraceAnomaly} proposed to employ a variational autoencoder with a Bayes model to detect anomalies reflected by latency.
\cite{MMTrace} extracted operation sequence and invocation latency from traces and fed them into a multi-modal LSTM to identify anomalies.
These anomaly detection methods rely on single-source data (i.e., traces) and ignore other informative data such as logs and KPIs.

Tremendous efforts~\cite{TBAC, NetMedic, MonitorRank, CloudRanger, MicroCause, AutoMAP, TraceRCA} have been devoted to root cause localization in microservice or service-oriented systems, most of which rely on traces only and leverage traditional or naive machine learning techniques.
For example, \cite{MEPFL} conducted manual feature engineering in trace logs to predict latent errors and identify the faulty microservice via a decision tree. 
\cite{MicroHECL} proposed a high-efficient approach that dynamically constructs a service call graph and ranks candidate root causes based on correlation analysis. 
A recent study~\cite{DyCause} designed a crowd-sourcing solution to resolve user-space diagnosis for microservice kernel failures. 
These methods work well when the latent features of microservices are readily comprehensible but may lack scalability to larger-scale microservice systems with more complex features.
Deep learning-based approaches explore meaningful features from historical data to avoid manual feature engineering. Though deep learning has not been applied to root cause localization as far as we know, some approaches incorporated it for performance debugging. 
For example, to handle traces, \cite{Seer} used convolution networks and LSTM, and \cite{Sage} leveraged causal Bayesian networks.

However, they rely on traces and ignore other data sources, such as logs and KPIs, that can also reflect the microservice status. 
Also, they either focus on anomaly detection or root cause localization leading to the disconnection in the two closely related tasks. The inaccurate results of naive anomaly detectors affect the effectiveness of downstream localization.
Moreover, many methods combine manual feature engineering with traditional algorithms, making it insufficiently practical in large-scale systems.
\section{Conclusion}
This paper first identifies two limitations of current troubleshooting approaches for microservices and aims to address them. The motivation is based on two observations: 1) the usefulness of logs and KPIs and the insufficiency of traces; 2) the unsatisfactory results delivered by current anomaly detectors.
To this end, we propose an end-to-end troubleshooting approach for microservices, \name, the first work to integrate anomaly detection and root cause localization based on multi-source monitoring data.
\name consists of powerful modality-specific models to learn intra-service behaviors from various data sources and a graph attention network to learn inter-service dependencies.
Extensive experiments on two datasets demonstrate the effectiveness of \name in both detection and localization. 
It achieves \textit{F1} of 0.988 and \textit{HR@1} of 0.982 on average, vastly outperforming all competitors, including derived multi-modal methods.
The ablation study further validates the contributions of the involved data sources.
Lastly, we release our code and data to facilitate future research.

\section*{Ackownledgement}
The work described in this paper was supported by the National Natural Science Foundation of China (No. 62202511), and the Research Grants Council of the Hong Kong Special Administrative Region, China (No. CUHK 14206921 of the General Research Fund).
\bibliographystyle{IEEEtran}
\bibliography{main}
\end{document}